\documentclass[aps,prd,onecolumn,preprint,groupedaddress,showpacs,nofootinbib]{revtex4}
\usepackage{amsmath, amsthm}
\usepackage{amsfonts}
\usepackage{graphicx}
\usepackage{dcolumn}
\usepackage{color}
\definecolor{darkred}{rgb}{.8,0,0}

\definecolor{darkblue}{rgb}{0,0,.7}

\definecolor{darkgreen}{rgb}{0,.8,0}

\def\ben{\begin{enumerate}} \def\een{\end{enumerate}}
\def\beq{\begin{equation}} \def\eeq{\end{equation}}
\def\beqn{\begin{equation*}} \def\eeqn{\end{equation*}}
\def\bea{\begin{eqnarray}} \def\eea{\end{eqnarray}}
\def\ba{\begin{array}} \def\ea{\end{array}}
\def\beann{\begin{eqnarray*}} \def\eeann{\end{eqnarray*}}
\def\beasn{\begin{sneqnarray}} \def\eeasn{\end{sneqnarray}}

\begin{document}
\title[]{Generalized G\"{o}del universes in higher dimensions and pure
Lovelock gravity}
\author{Naresh Dadhich $^{a,\;b}$}\email{nkd@iucaa.in}
\author{Alfred Molina{$^c$}}\email{alfred.molina@ub.edu }
\author{Josep M Pons{$^c$}}\email{pons@fqa.ub.edu}
\affiliation{$^{a}$ Center for Theoretical Physics, Jamia Millia Islamia, New Delhi~110025, India}
\affiliation{$^{b}$ Inter-University Center for Astronomy and Astrophysics, Post Bag 4, Pune 411 007,
India}
\affiliation{$^{c}$ Departament de F\'\i sica Qu\`antica i Astrof\'\i{s}ica and
Institut de Ci\`encies del Cosmos (ICCUB), Facultat de F\'\i sica, Universitat de Barcelona, Mart\'{\i} Franqu\`es 1,
E-08028 Barcelona, Catalonia, Spain.}

\begin{abstract}
G\"{o}del universe is a homogeneous rotating dust with negative $\Lambda$ which is a direct product
of three dimensional  pure rotation metric with a line. We would generalize it to higher
dimensions for Einstein and pure Lovelock gravity with only one $N$th order term.  For higher dimensional generalization,
we have to include more rotations in the metric, and hence we shall begin with
the corresponding pure rotation odd $(d=2n+1)$-dimensional metric involving $n$ rotations, which eventually can be extended by a direct product with a line or a space of constant curvature for yielding higher dimensional G\"{o}del universe. The considerations of $n$ rotations and also of constant curvature spaces is a new line of generalization and is being considered for the first time.
\end{abstract}


\maketitle

\section{Introduction}
The G\"{o}del universe \cite{godel} is one of the most interesting and at the same time enigmatic solutions of Einstein equation. It describes a homogeneous rotating dust in a stationary configuration with negative cosmological constant, $\Lambda$, in which  repulsion due to rotation is balanced by attraction due to negative $\Lambda$, and $\rho=-2\Lambda$. It is singularity free stationary cosmological solution. It is an exact solution of Einstein equation which is free of singularity and admits closed time-like curves. Both these features were
unexpected and rather bizarre. This is perhaps the reason for the large body of work on it as well as on the
related issue of time travel. For a rotating homogeneous dust, it is conceivable that linear velocity $r\omega$ corresponding to angular velocity $\omega$ could exceed velocity of light at some critical finite radius. This would then give rise to closed time-like curves (CTC), and it becomes visible if one transforms the metric into cylindrical coordinates. It is this feature that makes the powerful
singularity theorems inapplicable because it defies one of the assumptions of the theorems. It is stationary and requires negative $\Lambda$ for
countering repulsion due to rotation. Singularity free property was so attractive that  people like Raychaudhuri \cite{ray1, ray2} thought that it may be possible to win over these undesirable features for constructing a singularity free cosmological model. This of course did not happen and in the process he discovered the celebrated Raychaudhuri equation \cite{ray3}. Existence of CTC conflicts upfront with chronology protection. Unlike Kerr black hole where they are hidden behind a horizon, here they are naked and hence violate chronology protection. Despite all this, it continues to attract lot of attention from relativists as well as high energy theorists, more specifically string theorists of late (see a recent paper \cite{lfwl} and the  references given in there).

The focus of recent activity among high energy theorists is on embedding of G\"{o}del universe in string
theory \cite{bd, ht, tak, bdpr, bhh, lfwl}, effort to manage chronology violation \cite{bghv, israel} and putting in
black holes in \cite{gh, cglp, bbcg, wu, bc, pw, wp}. In classical GR as well, there is a huge literature on it, too vast
for comfort, yet we mention a few representative works. Inclusion of electromagnetic field as well as its
various properties have been studied in \cite{wp, bb, bz, rg, rt, rat}, and its analogues have also been found in Brans-Dicke \cite{anpps} and $f(R)$ gravity. On this background, it would not be out of place to study
G\"{o}del universe in higher dimensions and in pure Lovelock gravity. This is what we wish to do in this
paper.

It is clear that there can exist no solution for a static dust distribution
without $\Lambda$ because it would in general collapse under its own gravity in all
dimensions greater than three. In three dimension, Einstein gravity is kinematic
because Riemann curvature is entirely given in terms of Ricci tensor, and hence there can exist no
non-trivial vacuum solution, and Weyl curvature is always zero in three dimension. Further
gravitational potential goes as $1/r^{d-3}$ which turns constant for $d=3$, hence dust cannot gravitate to
collapse down and it can freely float as it is without any support. However in dimension, $d>3$,
it would gravitate and hence a positive $\Lambda$ is required to counter-balance gravitational attraction for
static equilibrium. This is the well-known Einstein universe. It is possible to define Lovelock Riemann,
$\mathbb{R}_{ijkl}$, which is a homogeneous polynomial in Riemann tensor, \cite{Dadhich:2008df} then the kinematic property, implying
$\mathbb{R}_{ij}=0 \Rightarrow \mathbb{R}_{ijkl}=0$, is in fact universalized to all critical odd
$d=2N+1$ dimensions in pure Lovelock gravity \cite{dgj, d16, c-d} where $N$ is degree of Lovelock
polynomial. By pure Lovelock we mean the action and consequently equation of motion has only one
$N$th order term without sum over lower orders. In the case of spherical dust distribution this
means that static dust in Einstein universe could freely float all by itself without collapsing in all critical odd $d=2N+1$
dimensions in pure Lovelock gravity, and requires no $\Lambda$ for support (see Appendix I). It becomes dynamic in
dimensions, $d>2N+1$, then a $\Lambda>0$ is required to counter-balance collapse. In particular,
in three dimension $\Lambda=0$ and dust is freely floating, to this when we add an angular coordinate
(from circle to sphere) to turn gravity dynamic and thereby switching on $\Lambda>0$, we arrive at the
famous Einstein universe in four dimension.

Note that Einstein universe metric is a product of constant curvature space with a time-like line and
hence it could be taken to higher dimensions simply by increasing dimension of constant curvature space. We have
further generalized it by taking product of Einstein universe with one or several constant curvature spaces. Interestingly the resulting space-time is a generalization of the
Einstein universe with density being supported by $\Lambda>0$. All this has been discussed in Appendix I.

The G\"{o}del metric \cite{godel} is a rotating analogue of Einstein universe with rotating dust.
It turns out that similar to Einstein universe its metric is also a product of three dimensional
pure rotation metric with a line describing a rotating dust distribution. Here there is repulsion due to
rotation which is countered by negative $\Lambda$. In the critical odd three dimension, gravity is kinematic,
hence matter does not gravitate and density then remains free and undetermined. To this when a flat coordinate -- line
is added, then density gets related, and the famous G\"{o}del universe
results in four dimension. Note that Riemann components corresponding to flat coordinate vanish and so do then
corresponding Ricci components. Now if the matter distribution has to
remain undisturbed, isotropy in spatial components will demand all spatial Ricci must vanish.
Then the remaining  equation will give the  relation $\rho+2\Lambda=0$. If one adds further lines
  -- flat coordinates, the solution remains undisturbed and any further extension in this way will
therefore be trivial. In the critical three dimension, density does not gravitate and hence it is negative $\Lambda$ that balances repulsion due to rotation. To this when a flat coordinate is added, which implies $\rho+2\Lambda=0$ and that is how density gets determined, while the balance between rotation and $\Lambda$ remains undisturbed. The solution is given in terms of rotation parameter alone. In three dimension, $\Lambda$ and rotation are related by the balance of force equation but solution involves both density and rotation, and hence it is not fully fixed by the equation of balance of forces. Only when flat coordinate is added, then it gets fully determined.   

Now if we wish to consider G\"odel metric in higher dimension, the only way it could be done is by adding more rotation parameters. That is, we begin with odd $d=2n+1$ dimensional  metric with $n$ rotations which we define as a pure rotation metric.
Here however gravity will be dynamic and hence density will participate in balance of force equation  because we have only one relation between the parameters density, rotations and $\Lambda$. When a flat extension is added a new relation between density and $\Lambda$ appears.

In higher dimensions, we should consider pure Lovelock gravity \cite{d16} as it universalizes kinematic property of gravity to all critical odd $d=2N+1$ dimensions. Secondly, bound orbits around a static object can exist in all dimensions $\geq 2N+2$ only in pure Lovelock gravity \cite{dgj1}. It is therefore pertinent to construct pure Lovelock analogues of  G\"{o}del universe. More importantly, does G\"{o}del solution in all $d=2N+1$ dimension have the similar behavior as that in three dimension where density does not gravitate and remains freely floating? When more than one rotations are involved, metric has very complicated structure and hence kinematic behavior may as well be not as transparently visible as in the case of one rotation in three dimension. It turns out that for the quadratic Gauss-Bonnet gravity, the solution in the critical five dimension does not bear out the kinematic behavior of density not gravitating, so much so that it admits $\Lambda=0$ without rotation vanishing. Then it is density that directly balances rotation in this case. However the solution cannot be extended to six dimension by addition of a flat coordinate because the additional required condition $\rho+4\Lambda=0$ admits no real solution.

The cubic $N=3$ case in the critical seven dimension admits two solutions, one of which accords to the expected kinematic behavior while the other does not. However both of them could be extended to eight dimensional G\"{o}del analogue. 

 It is indeed most intriguing and puzzling that kinematic behavior in the critical odd $d=2N+1$ dimension is not in general carried over to pure Lovelock gravity for $N=2, 3$. Note here that kinematicity property as Riemann being expressed in terms of Ricci has been established in general \cite{c-d} as a geometric property  which is true in general. In spherical symmetry it meant that that gravitational potential is constant in all critical odd $d=2N+1$ dimensions. This means that gravitational potentials for G\"odel metric involving more than one rotation do not become constant in the critical odd $d=2N+1$ dimension for $N>1$.
More importantly we do not have pure GB vacuum solution involving two rotations. That is why we really do not know how gravitational potentials behave.

The general procedure we shall follow in constructing higher dimensional G\"odel universe is as follows. We begin with a pure rotation metric in odd dimension and then take its direct product with a line or a space of constant curvature to get the corresponding G\"odel universe. The remarkable feature of this procedure is that addition of flat coordinate or constant curvature space does not disturb the pure rotation metric. It only implies a relation $\rho+2N\Lambda=0$ for flat extension or a relation involving $\rho, \Lambda$ and constant curvature of added space. The constant curvature does participate in balance of force equation with rotation, $\Lambda$ and density. It could be positive or negative depending upon the situation. It may also be noted that generalization of the G\"odel metric with more than one rotations for going to higher dimensions as well as product of pure rotation metric with constant curvature space is being considered for the first time.

The main aim of the paper is to generalize the rotating dust G\"{o}del universe to higher dimensional
space-times and to pure Lovelock gravity.
The paper is organized as follows: In the next section, we set up pure Lovelock equation of motion
for dust and $\Lambda$,
which will include Einstein gravity for $N=1$. Our aim is to study static
and stationary dust distribution with $\Lambda$. In sec III we introduce the standard G\"odel universe. In sec IV we discuss general features of pure rotation metrics that we will find in the following sections in Secs V-VIII as higher dimensional G\"odel generalizations in Einstein as well quadratic and cubic pure Lovelock gravity. Section IX is devoted to extended G\"odel metrics with number of 
rotations being less than the Lovelock order. Section X is concerned with particular rotation solutions in which all
the rotation parameters are equal and the constants $c$ are also equal. Appendix I gives some general
relations of Einstein spaces in Lovelock gravity, while in Appendix II we establish the conditions that follow when odd dimensional pure rotation 
solutions are extended by addition of a flat coordinate. Finally Appendix III contains some
details of seven dimensional pure GB solution. We end up with a discussion.

\section{Lovelock equation of motion}
\label{loveom}
The action of Lovelock gravity and the corresponding equation of motion are given by a sum of homogeneous polynomial terms in the Riemann curvature, each of them multiplied by a coupling $\alpha_k$ with length dimension $L^{2(k-1)}$ relative to the Einstein term. Action and equations for these theories are most simply written in terms of differential forms.\footnote{We are going to take the signature of the metrics as $-++\cdots+$}
\begin{eqnarray}
\mathcal{L}&=&\sum_{k=1}^N\alpha_k\,\frac{2^k\, \epsilon_{a_1 a_2\cdots a_d}}{(2k)!(d-2k)!}\,R^{a_1a_2}\wedge \cdots \wedge R^{a_{2k-1}a_{2k}}\wedge e^{a_{2k+1}}\wedge \cdots \wedge e^{a_d} ~,\\
G_{\ c}^{b}&=& \sum_{k=1}^N \alpha_k\,\frac{2^{k-1}\,\epsilon_{a_1 a_2\cdots a_{d-1} c}}{(2k)!(d-2k-1)!}\,\,R^{a_1a_2}\wedge \cdots \wedge R^{a_{2k-1}a_{2k}}\wedge e^{a_{2k+1}}\wedge \cdots \wedge e^{a_{d-1}}\wedge e^b ~.
\label{einstein}
\end{eqnarray}

Alternatively, one can introduce a set of $(2k,2k)$-rank tensors \cite{kastor} product of $k$ Riemann tensors, completely antisymmetric, both in its upper and lower indices,
\begin{equation}
\left.\right.^{(k)}\mathbb{R}^{b_1 b_2 \cdots b_{2k}}_{a_1 a_2 \cdots a_{2k}}= R^{[b_1 b_2}_{\quad \quad [a_1 a_2}\cdots R^{b_{2k-1} b_{2k}]}_{\qquad \qquad a_{2k-1} a_{2k}]}~.
\label{kastensor}
\end{equation}
With all indices lowered, this tensor is also symmetric under the exchange of both groups of indices, $a_i\leftrightarrow b_i$. In a similar way we will denote the contractions of $\mathbb{R}$ simply as
\begin{equation}
\left.\right.^{(k)}\!\mathbb{R}^{b_1 b_2 \cdots b_{J}}_{a_1 a_2 \cdots a_{J}}=\left.\right.^{(k)}\!\mathbb{R}^{b_1 b_2 \cdots b_{J} c_{J+1} \cdots c_{2k}}_{a_1 a_2 \cdots a_{J} c_{J+1} \cdots c_{2k}} \qquad ; \quad \forall \, J<2k ~.
\end{equation}
In terms of these new objects we can now write
\begin{equation}
\mathcal{L}=-\sum_{k=1}^N \alpha_k\, {}^{(k)}\mathbb{R} \quad \text{and} \quad
\mathbb{G}^a_{\ b}=\sum_{k=1}^N \alpha_k\left(k\,{}^{(k)}\mathbb{R}^a_{\ b}- \frac12 {}^{(k)}\mathbb{R}\,\delta^a_{\ b}\right)
\end{equation}
In any of the formulations, this reduces to the usual Einstein action $N=1$ and to Gauss-Bonnet(GB) for $N=2$.  Note that the $N$th order term in the action gives a non-trivial contribution to the equations only for $d\geq2N+1$.

For pure Lovelock, we have only one $\alpha_k$, $\alpha_k=\alpha_N, \alpha_j=0,j\neq N$ \cite{dgj,nkd},
and the equations of motion (EOM) are given by
\beq
\mathbb{G}^\mu_{\ \nu}\equiv N \mathbb{R}^\mu_{\ \nu} -\frac{1}{2}\mathbb{R}\,
\delta^\mu_{\ \nu}=T^\mu_{\ \nu} - \Lambda \delta^\mu_{\ \nu}\,.\label{Eeq}
\eeq
where the units are chosen such that the corresponding gravitational constant, $\alpha_N =1$.

We shall consider perfect fluid distribution as the source which is given by the energy-momentum tensor
\beq
T^\mu_{\ \nu} = (\rho +p) v^\mu v_\nu + p \,\delta ^\mu_{\ \nu} \,.
\label{emtensor}
\eeq
In particular we shall further specialize to dust for which $p=0$.

\section{G\"{o}del universe in Einstein gravity}
The famous G\"{o}del metric describes a singularity free stationary space-time filled with rotating
dust and negative cosmological constant. For equilibrium it is centrifugal repulsion that
balances attraction due to dust and negative $\Lambda$. The undesirable feature is that it admits
closed time-like curves. Despite this it is one of the most interesting and highly worked on
solutions of Einstein's equation. It is described by the metric \cite{Stephani:2003tm_2},
\beq
ds^2=\frac{2}{k} \left[-(dt+e^{x_1}dx_2)^2+(dx_1)^2+ \frac12 e^{2x_1}(dx_2)^2+(dx_3)^2 \right]
\eeq
and $k$,  the rotation parameter, the constant density and $\Lambda$ are related, $\rho=-2\Lambda=8k$.

Note that $x_3$ is a 'flat' coordinate. This means it is obtained by adding a flat coordinate to three
dimensional solution of a rotating dust with $\Lambda$. Any further addition of flat coordinates will
however be trivial.  G\"{o}del universe is therefore a direct product of three dimensional pure rotation metric with a line.

For constructing higher dimensional  G\"{o}del universe we would therefore employ the same method; i.e.
construct odd dimensional solution only with rotation spaces free of flat coordinate in the G\"{o}del metric ansatz, and then add flat coordinate to write next even dimensional solution.  In general we could add not only flat coordinate but also a space of constant curvature, and hence generalized G\"{o}del space-time would be a direct product of odd dimensional pure rotation space-time and constant curvature space, $M^{n_1}\times K^{n_2}$ where $d=n_1+n_2$ with $n_1$ odd. The flat space is included when constant curvature is zero.

\subsection{{Pure rotation three dimensional metric}}
\label{g3d}

Let us write the metric in the form
\beq
ds^2= -(dt + f(x_1) dx_2)^2 + dx_1^2  + g(x_1)^2 dx_2^2
\label{3dmetr}
\eeq

We shall work in the orthonormal basis:
$$\omega^0\equiv dt + f(x_1) dx_2,\,
\omega^1\equiv dx_1,\, \omega^2=g(x_1) dx_2,\,$$

The  equations of motion Eq. (\ref{Eeq}) imply the following conditions:

{\bf(a)} Vanishing of off diagonal components of $\mathbb{G}$ determine
function $g$ in terms of $f$,
\beq
 g= \frac{1}{\sqrt{ k}}f'\,,
\label{gequ}\eeq
where $k$ is a positive constant related (see section \ref{rot}) to dust rotation;

{\bf(b)} All diagonal components must be constant, which determines the functional form of
$f$ as
\beq
f(x_1)= A\, {\rm e}^{\sqrt{c}\, x_1}+
B\, {\rm e}^{-\sqrt{c}\, x_1}\, ,
\label{fequ}
\eeq
with $c$ a real constant. If fact $f$ admits also an arbitrary additive constant which
can be absorbed by a redefinition of time.

{\bf(c)} The isotropy condition implying equality of all spatial components, $\mathbb{G}^i_{\ i} =-\Lambda$, defines the equilibrium condition in terms of $k$ and $\Lambda$; and it is given by
\beq
4\Lambda + k =0.
\label{4lambdak}
\eeq
indicating repulsion due to rotation being balanced by negative $\Lambda$.
$k$ can be zero with an appropriate limit procedure explained in Section V.

{\bf(d)} Finally $\mathbb{G}^0_{\ 0}=T^0_{\ 0}- \Lambda \delta^0_{\ 0}= -\rho - \Lambda$ determines
the constant $c$ in Eq. (\ref{fequ})
in terms of $k$, $\Lambda$ and $\rho$ as given by
\beq
c= -\rho-4\Lambda = \frac{k}{2}-(2\Lambda+\rho).
\eeq
The latter expression for $c$ would be appropriate for future considerations.

Note that in the equilibrium condition only $k$ and $\Lambda$ are involved and it is free of density $\rho$ which could be freely chosen without affecting the equilibrium. This means that space-time is not uniquely defined unless density gets determined. That happens when a flat coordinate, i.e. a direct product with a line as it implies $\rho+2\Lambda=0$.

Once Eq. (\ref{4lambdak}) is satisfied, the density $\rho$ will determine the value of
the parameter $c$. Depending of the sign of $c$, the solutions for Eq. (\ref{fequ}) will be hyperbolic
functions or trigonometric. There is an infinity of metrics sharing the same equilibrium condition
Eq. (\ref{4lambdak}).

As a matter of fact, we should in principle have a solution with $k=\Lambda=0$ with density freely
floating without any support. This case is in fact obtainable from the
metric Eq. (\ref{3dmetr}) through the limiting procedure $k\to 0$ outlined in Sec. \ref{rot}.
The end result is Einstein universe, which has $\Lambda=0$ in three dimension.
This bears out the fact that gravity is kinematic (vanishing of Ricci implies Riemann zero,
hence no non-trivial vacuum solution) in three dimension, and density needs no support to stay put.

Also $\rho=0$ is admitted
with $c=k$ which means rotation and $\Lambda$ balancing each-other. In fact $\Lambda$ vanishes for Einstein universe in pure Lovelock gravity in all critical odd $d=2N+1$ dimensions,
as we show in Appendix I,  Eq. (\ref{lambdarhoEspaces2}).

\subsection{The G\"odel universe in four dimensions}

We now take the solution given by Eqns. (\ref{gequ}), (\ref{fequ}), to four dimensions by adding a flat
coordinate. As argued earlier, now all spatial Ricci components vanish which determines $\rho$ in
terms of $\Lambda$ as $\rho=-2\Lambda$ or $c=k/2$. This is a necessary and sufficient condition for
vanishing of spatial Ricci components to be zero. A general proof of this including the pure Lovelock
case is given in Appendix II. Now the solution takes the form
\beq
f(x_1)= A\, {\rm e}^{\sqrt{\frac{k}{2}} x_1}+
B\, {\rm e}^{-\sqrt{\frac{k}{2}} x_1}
\eeq
with $A$ and $B$ arbitrary constants.

With a change of variable $x'_1=x_1\sqrt{k/2}$, $B=0$ and corresponding changes in other
coordinates, the  metric then takes the standard form
\beq
ds^2=\frac{2}{k} \left[-(dt+e^{x_1}dx_2)^2+(dx_1)^2+ \frac12 e^{2x_1}(dx_2)^2+(dx_3)^2 \right]
\eeq

To this four dimensional G\"{o}del universe, one can add as many flat coordinates as one
wishes without disturbing its character; i.e. the solution continues to be the same on addition and hence any
further additions are trivial.

 \section{Pure rotation metrics in higher dimensions and in pure Lovelock \label{morehigher}}

Here we  consider some general features of the solutions to be  discussed later and they hold good  regardless of dimensionality and Lovelock order.  As mentioned earlier we can go to higher dimension in this metric ansatz by adding rotation parameters.
 For a given Lovelock order $N$,
in higher dimension, $d= 2n+1\geq 2N+1$, we shall have $n$ rotations, $k_a$, and $2n$ functions $f_a, g_a, a=1,..n$. Then pure rotation metric takes the form
\beq
ds^2= -(dt + \sum_{a=1}^n f_a(x_{2a-1}) dx_{2a})^2 +
\sum_{a=1}^n g_a(x_{2a-1})^2 dx_{2a}^2 + \sum_{a=1}^n dx_{2a-1}^2
=-\omega_0^2+\sum_{a=1}^n (\omega_{2a-1}^2+\omega_{2a}^2)\,
\label{Godel-n-EH-d= 2n+1}
\eeq
where
\beq
\omega_0\equiv dt +
\sum_{a=1}^n f_a(x_{2a-1})dx_{2a},\qquad  \omega_{2a-1}\equiv dx_{2a-1},\qquad  \omega_{2a}
\equiv g_a(x_{2a-1}) dx_{2 a}.
\eeq

It turns out that the relation between functions $g$ and $f$ and the form of solution for $f$
given by Eqns. (\ref{gequ}-\ref{fequ}) are simply carried over to higher dimensions as well as in
pure Lovelock case, and they read as follows:
\beq
g_a= \frac{1}{\sqrt{ k_a} }f'_a\,,
\label{rot-n-general1}
\eeq
where the constants $k_a$ must be positive to preserve the signature of the metric.

As the diagonal components of $\mathbb{G}$ must be constants then we obtain for the functions $f_a$
\beq
f_a(x)=
A_a {\rm e}^{\sqrt{c_a}\ x}+
B_a {\rm e}^{-\sqrt{c_a}\ x}
\label{rot-n-general2}
\eeq
and the constants $c_a$ must be constrained to satisfy the isotropy conditions (that is,
the equality of all the spatial components of $\mathbb{G}$).
The  diagonal spatial components $\mathbb{G}^{2a-1}_{2a-1}$ and $\mathbb{G}^{2a}_{2a}$  are
always equal,
which gives $n$ relations $\mathbb{G}^{2a}_{2a}=-\Lambda$ and solving these equations we
obtain in general the constants $c_a$ in terms of $k_a$ and $\Lambda$. Sometimes these $n$ equations
given by isotropy conditions are not independent, in which case one can not determine all the $c_a$
from them. These exceptional cases  result in a relation between $\Lambda$ and $k_a$, defining the equilibrium condition. This happens precisely in our previous three dimensional solution
and also it happens for one of the Lovelock N=3 solutions in Eq. (\ref{N3_1isot}).

The expressions for $\mathbb{G}^{2a}_{2a}$ are homogeneous polynomials in $c_a$ and $k_a$ of degree equal
to the Lovelock order $N$.
Finally the relation $\mathbb{G}^0_0=-\rho-\Lambda$ in general gives us a relation between density, cosmological
constant and the parameters $k_a$.

If all the parameters $k_a$ are equal and also the constants $c_a$ are equal the isotropy conditions
are automatically verified.

\section{Rotation}
\label{rot}
Throughout the paper we will consider for pure Lovelock gravity the ansatz
Eq. (\ref{Godel-n-EH-d= 2n+1})  and some generalizations with factor spaces of constant curvature
or flat coordinates. We define $d= 2n+1$ case as pure rotation G\"odel  space without any flat coordinate.
We shall however consider only cases N=1, 2, 3 but the results would be true for any Lovelock order $N$.

It is clear that the functions $f_a,\,g_a$ will always be related by
\beq
g_a= \frac{1}{\sqrt{ k_a} }f'_a\,,\ ( k_a>0)\label{rel_f_g}
\eeq
Then the velocity field $e_0 \equiv v^\mu\partial_\mu = \frac{\partial}{\partial\, t}$
will be shear and expansion free, and rotation will be given by
\begin{equation}
\Omega =-\frac12 \sum_{a=1}^{n} \sqrt{k_a}\, \omega_{2a-1} \wedge \omega_{2a}
\label{therot}
\end{equation}
That is, the parameters $k_a$ determine completely the rotation of the velocity field.

We may compute for $2$-form Eq. (\ref{therot}) the invariant scalar
\beq S= \Omega_{\mu\nu}\eta^{\mu\rho}\eta^{\nu\sigma} \Omega_{\rho\sigma} = \frac12\sum_{a=1}^n k_a \,,
\label{inv-rot}
\eeq
In general we can build more invariants by computing the traces of even
products of $\Omega$ and we can
see that all these traces can be written in terms of the $k_a$.
We conclude that the independent rotation
invariants are just the parameters $k_a$.  The number $n$ of such parameters identifies the
number of independent rotations, each involving a pair of coordinates $x_{2a-1},\,x_{2a}$.

This result depends only on the form of the metric Eq. (\ref{Godel-n-EH-d= 2n+1})
and of the relation Eq. (\ref{rel_f_g}), and has nothing to do with Lovelock order of gravity.

\subsubsection{The case $k= 0$}
\label{subsectk0}

Let us mention that we can cancel a rotation (for coordinates $x_{2a-1},\,x_{2a}$, say)
by sending the corresponding $k_a \to 0$. To proceed consistently we must first redefine coefficients
$\displaystyle A_a \to \sqrt{\frac{k_a}{c_a}} A_a,\
B_a \to -\sqrt{\frac{k_a}{c_a}} B_a$ and then take the limit $k_a \to 0$ on the metric coefficients.
This limit will cancel rotation for these coordinates $x_{2a-1},\,x_{2a}$ and instead a constant
curvature space for these two coordinates will appear. The limit will make 
$f_a=\rm{constant}$ (the additive constant that is always available) and
the function $g_a$ is then  determined as 
$\displaystyle  g_a(x_{2a-1}) = A_a {\rm e}^{\sqrt{c_a}\ x_{2a-1}}+
B_a {\rm e}^{-\sqrt{c_a}\ x_{2a-1}}$. Thus under the limit $k_a \to 0$, 
the space spanned by  $x_{2a-1},\,x_{2a}$ becomes a factor 
space of constant curvature, the curvature being $-c_a$. 

This is always the end result for this kind of
limits. Since factor spaces of constant curvature will be considered shortly as generalization of
Eq. (\ref{Godel-n-EH-d= 2n+1}), we will not dwell further on this issue of the limit $k_a \to 0$.

\section{G\"odel universe  in higher dimensions in
 Einstein gravity}

\subsection{Pure rotation G\"odel metrics in Einstein gravity \label{EH} }

We have the pure rotation metric Eq. (\ref{Godel-n-EH-d= 2n+1}) verifying Eqns. (\ref{rot-n-general1})
and (\ref{rot-n-general2}). In this case we obtain as isotropy conditions
the following $n$ linear  equations for $n>1$, (the case $n=1$ has been discussed in section \ref{g3d})
\beq
c_a=\frac{k_a}{2}-\frac{1}{n-1}\left(\frac{\sum_{b=1}^n k_b}{4}+\Lambda\right),\label{eq_19}
\eeq
which relates $\Lambda, k_a$ and $c_a$. Further invoking the remaining equation, $\mathbb{G}^0_{\ 0} = -\rho - \Lambda$, we get
\beq
\rho= \frac{1}{n-1}\left(\Lambda+\frac{2n-1}{4}\sum_{a=1}^n k_a\right),\label{eq_20}
\eeq
which displays how density and $\Lambda$ together counter balance the centrifugal repulsion due to rotation.
In terms of $\rho $ and $\Lambda$ we can also  write using Eq. (\ref{eq_20}) in Eq. (\ref{eq_19}) for $n=1,2\cdots$
\beq\  c_a= \frac{k_a}{2}-\frac{1}{2n-1}(2 \Lambda +\rho), \qquad
\sum_{a=1}^n k_a=\frac{4}{2n-1}((n-1)\rho-\Lambda)\,,\qquad  k_a>0\,, a= 1,...n\,.
 \label{EH-Godel-n-constr}
\eeq

Let us consider here some particular situations:

\noindent {\bf(a)} When $\Lambda=0$,
then $\sum_{b=1}^n k_b =4\rho(n-1)/(2n-1),\; c_a = k_a/2 - \sum_{b=1}^n k_b/(4(n-1))$ where density is
balancing rotation (also $\Lambda$ can be positive if $\rho>\Lambda/(n-1)$).

\noindent {\bf(b)} When $\rho=0$,
then $\sum_{b=1}^n k_b=-(4\Lambda)/(2n-1),\; c_a =  k_a/2+1/2 \sum_{b=1}^n k_b$.

Now when we add flat direction, which implies the condition, $\rho+2\Lambda=0$, (Appendix II)
we have
\beq
 c_a= \frac{k_a}{2}\, ,\qquad \sum_{a=1}^n k_a=-4\Lambda\,.
\eeq
This is the standard form for G\"odel universe. It is interesting that this form carries forward in higher dimensions.  That is when flat coordinate is added, there is direct relation between
$\sum k_a$ and $\Lambda$ and also $\rho$ and $\Lambda$ are separately related. Both
relations together exhibit the equilibrium of the configuration.

\subsection{Product of pure rotation metric with constant curvature space  in Einstein gravity}

Here we write the metric as product of pure rotation metric with space of constant curvature and it is given by
\beq
ds^2= -(dt + \sum_{a=1}^n f_a(x_{2a-1}) dx_{2a})^2 +
\sum_{a=1}^n g_a(x_{2a-1})^2 dx_{2a}^2 + \sum_{a=1}^n dx_{2a-1}^2+
\frac{1}{1-\sigma\, r^2} dr^2 + r^2 d\Omega_{(l-1)}^2\,
\label{Godel-n-EH+ctCurvature}
\eeq
where $l\geq2$, $d=2 n +1+l$ and $\sigma$ denotes constant curvature.

Now isotropy conditions and $\mathbb{G}^0_{\ 0} = -\rho - \Lambda$ would lead to
\beq
c_a= \frac{k_a}{2}-\frac{1}{d-2}(2 \Lambda +\rho), \qquad
\sum_{a=1}^n k_a=\frac{1}{d-2}(2(d-3)\rho-4 \Lambda)\,,\qquad  k_a>0\,, a= 1,...n\, .
\label{EH-Godel-n-withcurv}
\eeq
This is the generalization the relations Eq. (\ref{EH-Godel-n-constr}) obtained for $d=2n+1$. Further additional isotropy condition which includes constant curvature, $\sigma$ implies
\beq
\sigma = \frac{1}{(d-2)(l-1)}(2 \Lambda +\rho)\,.
\eeq

Let us consider the particular cases as done earlier for $\Lambda=0$, and $\rho=0$, as follows:

{\bf(a)} For $\Lambda=0$ we shall have
\beq
\sum_{a=1}^n k_a = 2\frac{d-3}{d-2} \rho
\eeq
and
\beq
 \rho = (d-2)(l-1)\sigma >0, .
\eeq
Note that here it is density that counterbalances rotation.

{\bf(b)} For $\rho=0$, it is
\beq
\sum_{a=1}^n k_a = \frac{-4\Lambda}{d-2} = -2(l-1) \sigma \, ,
\eeq
where $\Lambda<0$ for countering rotation.
Thus product with constant curvature space provides a richer structure to generalized G\"odel universe.

One can also consider introduction of several spaces of constant curvature, with the metric taking form
\bea ds^2&=& -(dt + \sum_{a=1}^n f_a(x_{2a-1}) dx_{2a})^2 +
\sum_{a=1}^n g_a(x_{2a-1})^2 dx_{2a}^2 + \sum_{a=1}^n dx_{2a-1}^2\nonumber \\
&+& \sum_{i=1}^s\left(
\frac{1}{1-\sigma_i\, r_i^2} dr_i^2 + r_i^2 d\Omega_{(l_i-1)}^2\right)\,,
\label{Godel-n-EH+morectCurvature}
\eea
with  $i=1,...s$. As usual we have $s$ new isotropy relations, one for each constant curvature space.
We have, in addition to Eq. (\ref{EH-Godel-n-withcurv}), the relations
\beq
\sigma_i = \frac{1}{(d-2)(l_i-1)}(2 \Lambda +\rho)\,.
\eeq
Again, when $2\Lambda +\rho=0$ all constant curvature spaces flatten out.

\section{G\"odel universe in pure Gauss-Bonnet gravity}\label{GBd5}


As before we start with the pure rotation metric in odd dimension, and then take its product with a line or space of
constant curvature to obtain the G\"odel universe analogue.   We take the metric in the form Eq.
(\ref{Godel-n-EH-d= 2n+1}) and the equations (\ref{rot-n-general1}) and (\ref{rot-n-general2}). We shall now refer to GB equation $\mathbb{G}^{(2)}_{ab}=T_{ab}$ (see Eq. (\ref{Eeq}) for $N=2$). We shall first consider the case of  the critical odd $d=2N+1=5$ dimension. The important point that  would emerge is that it cannot be extended further to six dimensional G\"{o}del solution by adding a flat coordinate. This happens because the condition for flat coordinate, $\rho+2N\Lambda=0$ does not have real solution for $N=2$. Thus there cannot exist a true pure
GB analogue of the G\"{o}del universe.

\subsection{Critical five dimensional}

Here pure rotation metric will have $n=2$ two rotations.
Using Eq. (\ref{Godel-n-EH-d= 2n+1}) now for $a=1,2$, we obtain for
the spatial diagonal components,
\beq
k_1(3k_2-4c_2)=-4\Lambda,\quad k_2(3k_1-4c_1)=-4\Lambda
\eeq

These equations give $c_a$ in terms of $k_b$ and $\Lambda$
$$c_a=\frac34 k_a+\frac{\Lambda}{k_b},\quad b\neq a$$
Again $c_a$ depend on $\Lambda$ and the constants $k$.

For the time component we obtain
\beq
-\frac{15}{4}k_1k_2+3(c_1k_2+c_2k_1)-4c_1c_2=-\rho-\Lambda .
\eeq
Substituting $c_a$ in terms of $k_a$ and $\Lambda$  we  can be write
\beq
\rho=\frac32 k_1\,k_2-\Lambda +4\frac{\Lambda^2}{k_1k_2}\,.
\label{kklambda2}
\eeq
This is the balance equation involving all three parameters, $\rho, \Lambda, k_a$. It is not a linear relation and hence it
reflects the very involved nature of the situation.  Notice that $\rho$ is always
positive and cannot be set to zero. Clearly density is not free here as it is balancing out rotation in the above equation (\ref{kklambda2}).
This is indeed contrary to the expectation in view of the kinematicity property for pure GB in the critical odd $d=5$ dimension, density should not have participated in the balance equation. It should have remained free as was the case for three dimension for Einstein gravity. What it perhaps indicates is the fact that for the metric with two rotations, GB kinematicity property is rather more involved. It is not always so transparent and
visible as the case for Einstein gravity. For
$\Lambda=0$ we have  $\displaystyle c_a=\frac{3}{4}k_a$, and  $\rho=\frac{3}{2} k_1k_2$
where density counterbalances rotations.

Note that depending on the values of $\Lambda$, $c_a$ can have both positive and negative values,
for instance if $\Lambda<-3k_1k_2/4$  then $c_a$ take negative values. In this case the solutions
\beq
f_a(x)= A_a e^{\sqrt{{c_a}}\,x}+ B_a
e^{-\sqrt{{c_a}}\,x}
\label{solGB-G2-5d}
\eeq
are trigonometric functions.

Now we should add a flat direction in order to extend the solution to get a pure GB analogue of the
G\"{o}del
solution. For that the required condition in general $\rho+2N\Lambda=0$ which for $N=2$ is
$\rho+4\Lambda=0$ (see Appendix II), equivalently $\mathbb G^0_0 +(2N-1)\mathbb G^1_1=0$. This has no
real solution because substitution $\displaystyle \rho\to -4\Lambda$ in Eq. (\ref{kklambda2})
yields complex results,
$$ k_1\,k_2= \Big(-1 \pm {\rm i} \sqrt{\frac{5}{3}}\Big) \Lambda\,.
$$
We conclude that, for pure GB gravity, the five dimensional solution cannot be extended to six dimensions
by adding a flat coordinate. Thus there exists no pure GB analogue of the G\"{o}del solution.

\subsubsection{Seven dimensional}
\label{GBd7}

As we have shown that the critical odd $5$-dimensional solution for pure GB cannot be extended to six
dimensional G\"odel universe. It turns out that the same continues to be true in higher dimensions as well.
Here we consider pure GB solution in seven dimensions with three rotations which again are not extensible
to eight dimensional G\"odel universe. For simplicity we shall consider a couple of particular solutions
while the general case is very involved and not very illuminating, though we give some details in Appendix III.

{\bf a)} Let's consider the case, for instance,  $c_1=m k_1,\, c_2=m k_2$, then the isotropy equation gives
\beq
m=\frac12-\frac{s}{4},\quad c_3=k_3\left(\frac{1}{4}+\frac{1}{s}\right) \, , \,
s\equiv k_3\left(\frac{1}{k_1}+\frac{1}{k_2}\right) \,
\eeq
and finally we obtain
\beq
4 \Lambda = 3{k_1}{k_2} (1-s)\, , \, \, \rho = \frac34 s k_1 k_2 \left(3+s\right)\,.
\eeq

Note that $s$ and $\rho$ are always positive and $\Lambda$ is positive for $s<1$ and negative for $s>1$.

On addition of a flat coordinate we must have $\rho+4\Lambda=0$ which requires
\beq
s^2-s+4=0 \, .
\eeq
Clearly it can have no real root and hence no extension to eight dimension for a G\"odel analogue.

{\bf b)} Another easy case is  of $c_a= m\, k_a$ for $a=1,2,3$, then the  isotropy
conditions imply all $k_a$ equal, $k_a=k$, and we have
$$c= \pm \frac{k}{2}\sqrt{\frac{\Lambda}{k^2}+\frac34},\quad \rho=k^2\left(12
+8\left(\frac{\Lambda}{k^2}+\frac34\right)\pm 9\sqrt{\frac{\Lambda}{k^2}+\frac34}\right)$$
Note that for density to be real $\Lambda>-3\,k^2/4$ and then $\rho$ is always positive.

Again the condition $\rho+4\Lambda=0$ for addition of a flat coordinate admits no real solution, and hence it cannot be extended to eight dimensions.

Thus it suggests that there can exist no pure GB analogue of the G\"odel universe.

\subsection{Product with constant curvature space in Gauss-Bonnet gravity}

As it happens with Einstein, the GB dynamics also allows for solutions as factors of G\"odel-like space-time
with space of constant curvature. Here we list some interesting solutions for the $7$-dimensional ansatz
\beq
ds^2= -(dt + f_1(x_1) dx_2+ f_2(x_3) dx_4)^2 + dx_1^2  + g_1(x_1)^2 dx_2^2 +dx_3^2 + g_2(x_3)^2 dx_4^2
+ \frac{dr^2}{1-\sigma\, r^2}  + r^2 d\phi^2\,,
\label{metric-GB-G2-2E-7d}
\eeq
with $f_a,\, g_a\ (a=1,2)$ given in Eq. (\ref{rot-n-general1})  and Eq. (\ref{rot-n-general2}). It describes the
product of a $5$-dimensional pure rotation metric with two rotations and a $2$-dimensional space of constant curvature.

\begin{itemize}
\item {\bf Solutions with $c_a = m k_a$}
\begin{enumerate}
\item For $m=\frac{1}{2}$ we have solutions with arbitrary $k_1, k_2$ and
$$
\Lambda= \frac{3}{4} k_1 k_2>0,\quad \rho=0,\quad \sigma =-\frac{k_1 k_2}{k_1+k_2}<0\, ,
$$
which written as
\beq
\frac43\frac{\Lambda}{k_1+k_2} +\sigma =0\,,
\eeq
shows that the negative constant curvature of the factor space counterbalances repulsion due to both
rotation and $\Lambda>0$.

\item For any $m$ we have solutions with $k_1=k_2=k$ and
$$\sigma= \frac{1}{8} k \left(-8 m-\frac{3}{m}+6\right),$$
$$\Lambda=k^2 \left(2 m (2 m-1)+\frac34\right),\ \
\rho= \frac{k^2 (4 m-3)^2 (2 m-1)}{4 m}\,.
$$

Note that for $m=\frac{3}{4}$ or $m=\frac{1}{2}$, $\rho$ vanishes. To have positive $\rho$ we need
$m\neq \frac{3}{4}$ and $m<0$ ($\Rightarrow \sigma>0$) or $m > \frac{1}{2}$ ($\Rightarrow \sigma<0$).
Here again
$\Lambda>0$ and hence repulsive, therefore density has to counterbalance rotation and $\Lambda$.
\end{enumerate}

\item {\bf Other solutions }

\begin{enumerate}
\item
For $c_1= {\displaystyle \frac{1}{4}} k_1$, $c_2= {\displaystyle\frac{1}{20}}(9 k_2 - 4 k_1)$ we have
solutions with
$$\sigma = - k_1,\; \Lambda=\frac{1}{2} k_1 k_2,\;  \rho=\frac{2}{5} k_1 (k_2- 6 k_1),
$$
which requires $k_2>6k_1$.
\item
In the special case $k_2= (7+4\sqrt{3})k_1$ we have solutions with
$c_1=c_2= \frac{1}{2} \sqrt{k_1 k_2}$,
$$ -4\Lambda=k_1 k_2 = \rho\,,\ \ \sigma=\frac{1}{4} \sqrt{k_1 k_2}.
$$
   \end{enumerate}

It is remarkable that these solutions satisfy $\rho + 4\Lambda=0$ the condition required for addition of a flat coordinate and therefore also admit an arbitrary number of additional flat directions. It is an interesting coincidence. Thus we see that whereas GB in $5$ dimensions with two rotations (parameters $k_1$, $k_2$) as discussed in section \ref{GBd5}
did not admit flat extensions,  now once this same
solution has been extended  to $7$ dimensions with a space of constant curvature,
the new space-time can be extended arbitrarily with flat directions.

 More general solutions are available, but not very illuminating, we skip them.
\end{itemize}


\section{G\"odel spaces in cubic Lovelock gravity}

\subsection{Pure rotation spaces in cubic Lovelock gravity}

In seven dimensions the metric will be as given in Eq. (\ref{Godel-n-EH-d= 2n+1}) with $n=3$, and the
solution has the same form  Eq. (\ref{rot-n-general1}) and Eq. (\ref{rot-n-general2}).
The isotropy equations $G^{2a}_{2a}=G^{2b}_{2b},\; b\neq a$ are
\beq
\frac{3}{8} k_3 (15 k_2 k_1- 12 k_1 c_2 -12  k_2 c_1 + 16 c_2c_1)=
\frac{3}{8} k_1(15 k_3 k_2 - 12 k_2 c_3 -12  k_3 c_2 + 16 c_3c_2)\,,
\label{N3_2isot0}
\eeq
\beq
\frac{3}{8} k_2( 15 k_1 k_3- 12 k_3 c_1 -12  k_1 c_3 + 16 c_1c_3)=
\frac{3}{8} k_1(15 k_3 k_2 - 12 k_2 c_3 -12  k_3 c_2 + 16 c_3c_2)\,,
\label{N3_2isot1}
\eeq
and have only the following two different solutions  (up to permutations of the indices)

\beq
\hspace{-3.3cm} {\bf (a)} \hspace{1.5cm}
c_1={\displaystyle\frac34} k_1,\quad c_2={\displaystyle\frac34} k_2, \label{N3_1isot}
\eeq
and
\beq
\hspace{-3.20cm} {\rm \bf (b)} \hspace{1.5cm}
c_a=m k_a\label{N3_2isot}
\eeq

Let us discuss both cases separately.

 {\bf Case (a)} We obtain from the isotropy condition,

\beq
4\Lambda+9 k_1 k_2k_3=0;\label{balanceN3}
\eeq
and $c_3$ is determined in terms of $\rho$ as
\beq
c_3=\left(2+\frac{\rho}{4\Lambda}\right)k_3\label{specialN3case}
\eeq
Notice that the isotropy condition is similar to the $3$-dimensional case in Einstein Gravity,
and so is density free and
arbitrary. It does therefore exhibit the kinematicity property for the cubic Lovelock gravity.
The limit of one $k$ going to zero is studied at the end of section IX.

More importantly the solution can be extended with addition of flat coordinate to the next even dimension $d=8$ by
imposing the condition $6\Lambda+\rho=0$, which simply determines $c_3=k_3/2$.

{\rm\bf Case (b)} In this case we have

\bea
&&\mathbb{G}^{(3)}{}^0_0=-\rho-\Lambda=-\frac38 k_1 k_2 k_3(64m^3-144m^2+180m-105),\nonumber \\
&&\mathbb{G}^{(3)}{}^i_i = -\Lambda=\frac38 k_1 k_2 k_3 (16 m^2-24 m+15),
\label{toshow}
\eea
and
\beq
\rho=-\mathbb{G}^{(3)}{}^0_0 + \mathbb{G}^{(3)}{}^i_i =-\frac32 k_1 k_2 k_3(16 m^3-40 m^2+ 51 m-30).
\eeq
Since the $k_a>0$, $\Lambda$ cannot vanish, in fact is always negative.

Here we cannot have the case of freely floating dust, and this case also defies kinematicity property.

Now to go over to next even dimension $d=8$ by adding a flat coordinate, we have the condition
$\rho+6\Lambda=0$ which translates to
\beq
32m^2(m-1)+15(2m - 1) = 0
\eeq
This is a cubic equation that always has one real root given by
\beq
m=\frac13\left(1+\frac14 \left( l^{1/3}-29 l^{-1/3}\right)\right)\label{Godelmvalue}
\eeq
where $l\equiv 199+9\sqrt{790}$. Thus the critical odd dimensional solution is extensible for $N=3$ by
adding a flat coordinate giving an analogue of G\"{o}del universe in eight dimension for $N=3$ pure
Lovelock gravity even though it does not obey the kinematicity property.

This suggests that pure Lovelock for $N=3$ analogues of the  classical G\"{o}del universe could always
be constructed by adding a flat coordinate to the critical odd dimensional solution.
Though we have only considered $N=2$ where the solution is inextensible while it is
extensible for $N=3$, this feature should hold good for all even and odd $N$.
\subsection{Product spaces in cubic Lovelock gravity  }

Thus higher dimensional analogues of the G\"{o}del universe would involve $n>1$ rotations.
Here we consider G\"odel universes which metrics are given as products of the standard rotation metrics
discussed before and spaces of constant curvature. To be specific we will add to the metric
Eq. (\ref{Godel-n-EH-d= 2n+1}) new $l=2,3$ dimensions spanning a space of constant curvature.
\beq ds^2= -(dt + \sum_{a=1}^3 f_a(x_{2a-1}) dx_{2a})^2 +
\sum_{a=1}^3 g_a(x_{2a-1})^2 dx_{2a}^2 + \sum_{a=1}^3 dx_{2a-1}^2+
\frac{1}{1-\sigma\, r^2} dr^2 + r^2 d\Omega_{(l-1)}^2\,.
\label{Godel-n-EH+ctCurvature1}
\eeq
So the dimension is $d=2 n +1+l$. With the same ansatz Eqns. (\ref{rot-n-general1}-\ref{rot-n-general2}) we obtain solutions but now the isotropy conditions requires that $k_1=k_2=k_3=k$ for the solution
$c_i=mk_i$, and $k_1=k_2=k$ for the solution $c_1=3/4\,k_1,\,c_2=3/4\,k_2,\,  c_3=k/2+k_3/4$
\begin{itemize}
\item $l=2$

{\bf a)} For $c=m k$ $$\sigma=\frac{k}{2}\frac{32(m^3-m^2)+15(2m-1)}{3-16m^2}$$
and $\Lambda$ and $\rho$ some involved polynomials of $m$. For $m<\sqrt{3}/4$  we have $\rho>0$, $\sigma>0$ and $\Lambda>0$.

{\bf b)} For $c_1=c_2=3/4\,k,\ c_3=k/2+k_3/4$
\beq
\Lambda=-\frac92 k^3,\quad \rho=9k\left(k+\frac{k_3}{2}\right)\left(k_3-\frac{k}{2}\right),\quad \sigma=
\frac12\left(\frac{k_3}{2}-k\right).
\eeq
Notice that $\Lambda<0$ and for $k_3>k/2$, $\rho>0$ and then $\sigma$ can be positive or negative. The case $k_3=2k$ implies vanishing $\sigma$ and indeed it is the case which allows for
the extension with flat coordinates, because then $\rho + 6 \Lambda=0$. For $k_3>2k$ it is $\sigma>0$.

\item $l=3$

{\bf a)} For $c=m k$ $$\sigma=\frac{k}{12}\frac{32(m^3-m^2)+15(2m-1)}{3-4m}$$
and $\Lambda$ and $\rho$ some involved functions of $m$. $\rho>0$ for $0.605<m<3/4$ in this interval
$\Lambda$ and $\sigma$ can be positive and negative.
We have $\sigma=0$ flat dimensions for the value
(\ref{Godelmvalue}) as it will be.

{\bf b)} $c_1=c_2=3/4\,k,\ c_3=k/2+k_3/4$, $$\sigma=-\frac{3k}{8}\left(1+\frac{k}{k-k_3}\right)$$
\beq
\Lambda=-\frac{9 k^2}{8}\frac{18k^2-2k_3^2-7kk_3}{k-k_3},\quad \rho=
\frac{9 k^2}{4}\frac{2k^2-2k_3^2+9kk_3}{k_3-k}.
\eeq
$\rho>0$ for $\displaystyle\frac{\sqrt{97}}{4}-9<\frac{k}{k_3}<1$.

To have flat dimensions $\sigma=0\Rightarrow k_3=2k$ i.e. $c_3=k_3/2$,
as it must be.
\end{itemize}

\section{Extended G\"odel metrics with $n<N$}Thus higher dimensional analogues of the G\"{o}del universe would involve $n>1$ rotations.
 Let us point out that  if we extend some pure  $n$-rotation G\"odel
spaces Eq. (\ref{Godel-n-EH-d= 2n+1})
with flat directions to get a space of
dimensions larger than $2n+1$, the tensor $\mathbb{G}^{\!{}^{(N)}}$ in the left hand side of Eq.
(\ref{Eeq}) vanishes identically for $N>n$, irrespective of any specification of the
functions $f_a,\,g_a$.

However we can extend the pure  $n$-rotation G\"odel spaces with constant curvature spaces in this
case we obtain non trivial solutions.

Here are some examples with one or two rotations admitting non-trivial constant curvature extensions
(dimensions $l_i$), always for metrics of the form Eq. (\ref{Godel-n-EH+morectCurvature}):

\underline{\bf For N=2, (GB)}

\underline{\bf $d=6$,\ $l=3$.}

There is a solution with $\rho=\Lambda=0$,  $c=k$ and $k$,\ $\sigma$ arbitrary.

\vspace{4mm}

\underline{\bf $d=7$,\ $l=4$.}

With $k$, $\sigma$ arbitrary, $\displaystyle c=\frac{3}{4} k - \sigma$,\
$\rho= 6\,\sigma (k + 4\, \sigma)$,\ $\Lambda= 6\,\sigma (-k + 2 \,\sigma)$.

In the particular case $\displaystyle \sigma= \frac{1}{4} k$ this solution satisfies $\rho+4\,\Lambda=0$
signaling (Appendix II) that it can be arbitrarily extended with additional flat coordinates.
In this case $\displaystyle c= \frac{1}{2} k$,\ $\rho=3\,k^2$,\
$\displaystyle \Lambda=-\frac{3}{4} k^2$.

\vspace{4mm}

\underline{\bf $d=7$,\ $l_1=2$,\ $l_2=2$.}

There is a solution with $c={\displaystyle\frac{k}{4}}$, $\Lambda=0$,\ $\rho=12\,\sigma_1\sigma_2$,\
$\displaystyle k=4\,\frac{\sigma_1\sigma_2}{\sigma_1+\sigma_2}$.
Notice that $\sigma_i>0$.

\underline{\bf For N=3} we have also solutions with $n<N$, for instance  $d=7$ with two rotations $k_1,k_2$  and a constant curvature space, $\sigma$ of dimension 2, we have
$$c_1=\frac34 k_1, c_2=\frac34 k_2$$ gives  a solution $$\Lambda=0\quad {\rm and}\quad  \rho=9k_1k_2\sigma >0 \, .$$
Here $\sigma$ must be positive.

\section{Some particular metrics with all $k_a$ equal}
The general metric Eq. (\ref{Godel-n-EH+ctCurvature}) with the functions verifying  Eq. (\ref{rot-n-general1})
and Eq. (\ref{rot-n-general2}) if all the
$k_i$ and all the $c_i$ are equal we can distinguish two cases if $c$ is positive and if  it is negative in the first case the functions $f_i$ are exponential functions, and in the second one are trigonometric functions in both cases we can write  $m\equiv |c|/k$ where $m>0$ and can be write in a coordinate system the metric with a form closer to the usual G\"odel form of the metric. In the general metric Eq. (\ref{Godel-n-EH+ctCurvature})
we can perform the following change of variable $x'_{2a-1}=\sqrt{|c|}\,x_{2a-1} $, and taking $B_a=0$ then we have the following
expression for the functions
$$f_a(x')=
A_a {\rm e}^{x'},\;{\rm or}\; A_a \sin{x'} \quad g_a(x')=\sqrt{m} f_a'(x')
$$
and changing also all the other coordinates $x'_{2a}=A_a\sqrt{|c|}\,x_i$ and also the radial coordinate for the
constant curvature space $r'=\sqrt{|c|}\,r$ and for the temporal coordinate $t'=\sqrt{|c|}\,t$.  If $c>0$ we get for the metric  Eq. (\ref{Godel-n-EH+ctCurvature}) the following expression
\begin{eqnarray}
ds^2= && \frac{1}{c}\bigg[-\bigg(dt + \sum_{a=1}^n e^{x'_{2a-1}} dx'_{2a}\bigg)^2
+\frac1m\sum_{a=1}^n e^{2x'_{2a-1}} dx_{2a}^2 + \sum_{a=1}^n dx'_{2a-1}{}^2+ \nonumber \\[1ex]
&& \frac{1}{1-\sigma'\, r'^2} dr'^2 + r'^2 d\Omega_{(l-1)}^2\bigg]
\end{eqnarray}
and if $c<0$ the following one
\begin{eqnarray}
ds^2= && \frac{1}{|c|}\bigg[-\bigg(dt + \sum_{a=1}^n \sin{x'_{2a-1}} dx'_{2a}\bigg)^2
+\frac1m\sum_{a=1}^n \cos^2{x'_{2a-1}} dx_{2a}^2 + \sum_{a=1}^n dx'_{2a-1}{}^2+ \nonumber \\[1ex]
&& \frac{1}{1-\sigma'\, r'^2} dr'^2 + r'^2 d\Omega_{(l-1)}^2\bigg]
\end{eqnarray}
where in both cases  $$\sigma'=\frac{\sigma}{|c|}$$

\section{Discussion and conclusions}\label{Discussion}
Let us summarize our main results. Throughout the paper
we rely on a generic pure rotation ansatz Eq. (\ref{Godel-n-EH-d= 2n+1}) and additions of constant curvature spaces Eq. (\ref{Godel-n-EH+morectCurvature}). If we extend a pure  $n$-rotation G\"odel space-time with flat directions to get a space of
dimensions larger than $2n+1$, the tensor in the left hand side of the EOM  Eq. (\ref{Eeq}), $\mathbb{G}^{N}$ with $N>n$,  vanishes identically irrespective of any specification of the functions $f_a,\,g_a$.
However we can obtain non trivial solutions with $N>n$ by extending the pure  $n$-rotation G\"odel spaces with additional constant curvature spaces.

The main body of the paper is devoted to the case $n\geq N$.
An universal feature of our solutions is that under the general ansatz
Eq. (\ref{Godel-n-EH-d= 2n+1}) or, more generally, Eq. (\ref{Godel-n-EH+morectCurvature}), we always obtain the
relations $\displaystyle g_a= \frac{1}{\sqrt{ k_a} }f'_a$ and the functions $f_a$ are always of the form
$\displaystyle f_a(x)= A_a {\rm e}^{\sqrt{c_a}\ x}+B_a {\rm e}^{-\sqrt{c_a}\ x}$.
Thus in the $N$-Lovelock dynamics,
the EOM Eq. (\ref{Eeq}) always become algebraic equations with a common structure: homogeneous polynomials
of degree $N$ in the parameters of the metric, $k_a,\, c_a,\, \sigma_i$, on the left hand side,
and linear terms in the parameters $\rho$ and $\Lambda$ on the right hand side. In general from these equation we get constraints among these parameters and balance equations relating  $k_a,\, \sigma_i,\, \rho,\, \Lambda$.

The rotation properties of these space-times are completely described by the parameters $k_a$. It is worth
mentioning that our solutions admit the limit for one or more of these parameters to vanish, in which case,
for each vanishing $k_a$ we obtain a two dimensional constant curvature space, and the associated parameter
$c_a$ becomes  the curvature of this space.

In the framework of Einstein gravity, the  G\"odel universe can be described as a direct product of a three dimensional pure rotation
space-time with a line --  flat one-dimensional space.   It is in fact a rotating analogue of
the Einstein universe which describes a constant density static sphere supported by a positive
$\Lambda$. Here it is rotating constant density stationary dust distribution supported by a negative
$\Lambda$. The latter balances repulsion due to rotation, and is related to density by the relation
$\rho+2\Lambda=0$. It is worth noting that density
does not directly counter rotation, it is done via negative $\Lambda$.  Einstein and G\"odel universes
are the two constant density distributions which require $\Lambda$ for support with opposite sign,
positive for the former while negative for the latter.

The general construction of G\"odel universe is that we begin with a pure rotation metric (9) in three
dimension which solves for constant density with $\Lambda$ to give Eqs (10-11). For getting to G\"odel
universe, we take a direct product of this metric with a line, which does not disturb the solution in
Eqs (10-11), and it only implies the relation $\rho+2\Lambda=0$. Any further addition of flat
coordinates would have no effect on the solution and hence would be trivial. This means that to get new non trivial solutions in higher dimensions one  would have to introduce more rotation parameters or constant curvature spaces in the metric as given in Eqns. (\ref{Godel-n-EH-d= 2n+1}), (\ref{Godel-n-EH+ctCurvature})
and the corresponding solution is given in Eqns. (\ref{rot-n-general1}), (\ref{rot-n-general2}). For an $n$ rotation parameters, the odd dimensional pure rotation space-time would be of dimension $d=2n+1$. For equilibrium,
repulsion due to rotation is to be counter-balanced by attraction due to negative $\Lambda$ and
density, and there would be a balance of force equation involving in general the rotation parameters, $\rho$ and $\Lambda$. When a flat coordinate is added to
the metric we get the new condition $\rho+2\Lambda=0$.   As before the solution in Eqns. (\ref{rot-n-general1}), (\ref{rot-n-general2}) is not disturbed by this
extension, and the balance of force equation would only involve rotation parameters and $\Lambda$.
This is how G\"odel analogue is constructed in higher dimensions for Einstein gravity.

However something special happens in three dimension where the balance of force equation is
$k=-4\Lambda$ ($k$ is the rotation parameter), and density  does not participate in the balance of forces. Here gravity is kinematic because Weyl curvature is zero implying Riemann being entirely defined in terms of Ricci. Thus there can exist no non-trivial vacuum solution, and hence dust does not gravitate, and needs no support to stay put as it is (this also happens for the spherical case as well, where Einstein universe in three dimension has freely floating arbitrary density without any $\Lambda$ support as shown in Appendix I). When flat coordinate is added then density gets roped in by the condition,  $\rho+2\Lambda=0$, and the previous balance of force remains undisturbed. We obtain the G\"odel universe. 

The question arises whether this feature is carried forward to pure Lovelock
where gravity is kinematic in all critical odd $d=2N+1$ dimensions (i.e.  Lovelock Riemann is given
in terms of the corresponding Ricci \cite{c-d})?  It turns out as shown in Appendix I that this feature is indeed carried forward for spherically symmetric Einstein universe or products of constant curvature spaces for pure Lovelock gravity. That is, in all critical odd $d=2N+1$ dimensions dust does not gravitate, as was the case in three dimension, and hence needs no support from $\Lambda$ to stay freely floating. 

The next question arises: Is it also true for the G\"odel system, particularly because of complexity
caused by more than one rotations and highly non-linear nature of Lovelock equation. Though gravity is kinematic in critical $d=2N+1$ dimensions, and hence one would expect dust to freely float here as well as it did in three dimension for $N=1$. It is most intriguing and puzzling that this does not happen in general for rotating dust pure Lovelock G\"odel solutions. It turns out that for pure GB, density does participate in balance of force equation which 
is now non-linear in rotations. Not only that it admits $\Lambda=0$ solution where density directly balances rotation.   However when flat coordinate is added, then the required condition
$\rho+4\Lambda=0$ admits no real solution. Hence there cannot exist a pure GB analogue of G\"odel
universe. In the case of cubic Lovelock $N=3$, there exist two solutions, one of which is in 
accordance with the expectation that in seven dimension density does not participate in the balance equation while for the other it does. 
In this case both solutions are interestingly extendable by addition of a flat coordinate to
eight dimensional G\"odel universe. In general, therefore G\"odel system  does not accord to the behavior of freely floating density in  three dimension ($N=1$) in all critical odd $d=2N+1$ dimensions.

What could be the reason for this strange and puzzling behavior? In spherical symmetry, pure Lovelock gravitational
potential goes as $1/r^\alpha\, \, , \alpha=(d-2N-1)/N$ which turns constant for $d=2N+1$. That is why
density does not gravitate. For G\"odel metric in higher dimensions involving more than one rotation, added to it highly non-linear nature of Lovelock equation makes the situation very involved. As a matter of fact we do not have analogue of Kerr solution in pure GB and hence it is difficult to decipher how would potentials behave in this case. This is a good indication of the complexity of the situation. What our study of $N=2, 3$ G\"odel solutions seems to indicate is that potentials do not in general turn constant in critical odd  $d=2N+1$ dimensions. Further this is also suggestive that for even $N$ there may exist no pure Lovelock G\"odel solution satisfying the relation $2N\Lambda+\rho=0$ while for odd $N$, it would always exist. 

All in all it is indeed the most surprising and intriguing result which we do not fully understand. It asks for a deeper and detailed analysis to clearly decipher what is really happening. We leave this for a future study.


\section*{ Appendix I. Einstein spacetimes in Lovelock gravity\label{I}}
A $d$-dimensional Einstein space-time is described by the metric
\beq
ds^2= -dt^2 +
\frac{1}{1-\sigma\, r^2} dr^2 + r^2 d\Omega_{(d-2)}^2\,.
\label{Einsteinspace}
\eeq
These metrics are solutions to the EOM Eq. (\ref{Eeq}) for a perfect fluid Eq. (\ref{emtensor}).
In the pressure-less case one can find the relation between the curvature $\sigma$ and the
cosmological constant $\Lambda$ or the fluid density $\rho$. In the general  $N$-order Lovelock
dynamics one has ($d\geq 2N+1$)
\beq
\Lambda_N = \frac12 \frac{(d-2)!}{(d-2N-2)!} \sigma^N \,,\qquad
\rho_N = N  \frac{(d-2)!}{(d-2N-1)!} \sigma^N\,,
\label{lambdarhoEspaces}
\eeq
out of which we obtain the relation between $\Lambda_N$ and $\rho_N$ (balance equation),
\beq
2 N \Lambda_N = (d-2N-1) \rho_N
\label{lambdarhoEspaces2}
\eeq

In terms of Lovelock scalar curvatures, for $N=1$ we have the
Riemann scalar curvature which is $\displaystyle R=\frac{(d-1)!}{(d-3)!}\sigma$. It generalizes to the
$N$-order Lovelock scalar curvature
(see section \ref{loveom}), with notation
${}^{(N)}\mathbb{R}\equiv \mathbb{R}_N$, as
\beq
\mathbb{R}_N= \frac{(d-1)!}{(d-2N-1)!}\sigma^N\,.
\label{lovelockcurvature}
\eeq
Thus
\beq\Lambda_N = \frac12\, \Big(\frac{d-2N-1}{d-1}\Big) \mathbb{R}_N\,,\qquad
\rho_N=\frac{N}{d-1} \mathbb{R}_N\,.
\label{lovelockcurvature-R}
\eeq

Einstein space-times can be generalized with products
of spaces of constant curvature. The metric (with dimension $d= \sum_{i=1}^s l_i + 1$) is
 \beq
ds^2= -dt^2 +\sum_{i=1}^s\left(
\frac{1}{1-\sigma_i\, r_i^2} dr_i^2 + r_i^2 d\Omega_{(l_i-1)}^2\right)\,,
\label{ProdEinsteinspaces}
\eeq
and the basic equations Eq. (\ref{lambdarhoEspaces2}) and Eq. (\ref{lovelockcurvature-R}) still hold.

Note that in the critical dimension $d=2N+1$ we have $\Lambda_N =0$ and
$\displaystyle \rho_N = \frac12 \mathbb{R}_N$,
thus explicitating for this static space-time the kinematic character of gravity in the critical dimension.
Indeed, we may have
static matter (described by the fluid density $\rho_N$) but it does not gravitate and hence it
does not need to be balanced by the compensating mechanism of a repulsive background with a
positive $\Lambda$, as it is the case for $d>2N+1$.

\section*{Appendix II } \label{II}
Here we use the tetrade basis. We take as the $0$-vector
of the tetrade the fluid velocity $e^0 =v^\mu\partial_\mu = \frac{\partial}{\partial\, t}$.

From now on indices in this appendix will refer always to flat indices. We will show
that in the case of a perfect fluid a necessary condition for a given solution of Eq. (\ref{Eeq})
for $N=1$ in some dimensionality to be enlarged to higher dimensions by the addition of flat
directions is that $\rho +2 \Lambda=0$.

Suppose that the flat direction is $z$ so the component $R^z_{\ z}$ of the Ricci tensor vanishes
$R^z_{\ z}=0$. On the other hand, the curvature scalar is $R = R^0_{\ 0} + R^i_{\ i} + R^z_{\ z}$.
and the Einstein equation Eq. (\ref{Eeq}) read
$G^\mu_{\ \nu}=R^\mu_{\ \nu} -\frac{1}{2} R\, \delta^\mu_{\ \nu}= T^\mu_{\ \nu}\,,$
with $\displaystyle T^\mu_{\ \nu}= Diag\{-\rho-\Lambda,-\Lambda,-\Lambda,...\}$.
Thus the non-diagonal components of $\displaystyle R^\mu_{\ \nu}$ must vanish and
$\displaystyle R^i_{\ i}=R^z_{\ z}=0$. This means that $R=R^0_{\ 0}$ and we get
$$G^0_{\ 0} = R^0_{\ 0} -\frac{1}{2} R^0_{\ 0} = \frac{1}{2} R^0_{\ 0} =
-\rho-\Lambda\,\qquad
G^i_{\ i}= -\frac{1}{2} R^0_{\ 0}= -\Lambda
$$
thus inferring $G^0_{\ 0}=-G^i_{\ i}$ and hence $-\rho-\Lambda =\Lambda$,
which is the condition we were looking for,
$$\rho+2\Lambda=0\,.
$$

This argument generalizes to Lovelock Lagrangians of order $N$. In this case we have
$$\mathbb{G}^\mu_{\ \nu}=N\mathbb{R}^\mu_{\ \nu} -\frac{1}{2}\mathbb{R}\, \delta^\mu_{\ \nu}\,.$$
The non-diagonal components of $\displaystyle \mathbb{R}^\mu_{\ \nu}$ must vanish and if there are
flat directions i.e $z$ then $\mathbb{R}^z_{\ z}=0$
and since all the spatial diagonal terms must coincide the non flat directions must vanish as well,
i.e. $\displaystyle \mathbb{R}^i_{\ i}=\mathbb{R}^z_{\ z}=0$. This means that
$\mathbb{R}=\mathbb{R}^0_{\ 0}$ and we get
$$\mathbb{G}^0_{\ 0} = N \mathbb{R}^0_{\ 0} -\frac{1}{2} \mathbb{R}^0_{\ 0} =
\frac{2N-1}{2} \mathbb{R}^0_{\ 0} = -\rho-\Lambda\,,\qquad
\mathbb{G}^i_{\ i}= -\frac{1}{2} \mathbb{R}^0_{\ 0}= -\Lambda
$$
that is $$\mathbb{G}^0_{\ 0} =-(2N-1)\mathbb{G}^i_{\ i}\,,$$
which implies
$$
\rho+2N\Lambda=0\,.
$$

This is the new condition analogous to the G\"{o}del condition $\rho+2\Lambda=0$
which holds in the Einstein case.

\section*{Appendix III: $7$-dimensional space-time in GB gravity\label{III}}
In $7$ dimensions we can examine the ansatz Eq. (\ref{Godel-n-EH-d= 2n+1}) for the metric with $n=3$.

We use the notation of Eq. (\ref{rot-n-general1}) and Eq. (\ref{rot-n-general2}) that is,
\beq
g_a=  \frac{1}{\sqrt{k_a}} f'_a\,,\ k_a>0, \qquad
f_a(x)= A_a e^{\sqrt{c_a}\,x}+ B_a e^{-\sqrt{c_a}\,x}\,,\, a=1,2,3\,,
\label{solGB-G3-7d}
\eeq
where $k_a>0$, $c_a$ are real and $A_a$ and $B_a$ are arbitrary. The independent algebraic equations
satisfied by
$k_1,k_2,k_3, c_1,c_2,c_3,\Lambda,\rho, $
become three equations quadratic in $c, k$ and linear in $\Lambda$ (coming form the isotropy
conditions) and the equation $\mathbb{G}^0_{\ 0}= T^0_{\ 0}$ which is quadratic in $c, k$
(but with no $c_i c_j$ terms) and linear in
 $\Lambda$ and $\rho$. One can manage to arrange three linear combinations of these equations which
 eliminate the quadratic terms in $c$. All in all the system becomes a linear system for the $c$ variables
\beq
M \left(\begin{array}{c} c_1\\ c_2 \\ c_3 \end{array}\right)+
\left(\begin{array}{c}-\frac{3}{4} ({k_1}-{k_2}) (4 \Lambda +{k_1} ({k_2}+{k_3})+{k_2} {k_3})\\
-\frac{3}{4}({k_1}-{k_3}) (4 \Lambda +{k_1} ({k_2}+{k_3})+{k_2} {k_3})\\
3 (4 \Lambda +9 {k_1}({k_2}+{k_3})+9 {k_2} {k_3}-2 \rho )\end{array}\right)=
\left(\begin{array}{c} 0\\ 0 \\ 0 \end{array}\right)
\label{linearsystem}
\eeq
where $M$ is
\beq
\left(
\begin{array}{ccc}
 - 3{k_2} {k_3}+ 3{k_1} ({k_2}+{k_3})+
 4 \Lambda  & -3 {k_2} {k_3}+3 {k_1}
   ({k_3}-{k_2})-4 \Lambda  & 0 \\
 - 3{k_2} {k_3}+3 {k_1} ({k_2}+{k_3})+4 \Lambda  & 0 &
  3{k_1} {k_2}-
   3({k_1}+{k_2}) {k_3}-4 \Lambda  \\
 -18 ({k_2}+{k_3}) & -18 ({k_1}+{k_3}) & -18 ({k_1}+{k_2}) \\
\end{array}
\right)
   \label{matrixM}
\eeq
 (We get solutions for $c_1,c_2,c_3$ as long as $Det(M)\neq 0$) together with a remaining equation linking
$k_1,k_2,k_3,\Lambda,\rho$.

The solutions for  $c_1,c_2,c_3$ of the system above are not very illuminating and we skip them.
The remaining equation has the form
 \beq
 \alpha_0 + \alpha_1\,\rho+ \alpha_2\,\rho^2=0\,.
 \label{remainigcondition}\eeq
 It is worth to remark a neat factorization in $\alpha_2$,
 \bea
    \alpha_2&=& \Big(-4 \Lambda -3 {k_3} ({k_1}+{k_2})+3 {k_1} {k_2} \Big)  \Big(4 \Lambda +3 {k_1}
   ({k_2}-{k_3})+3 {k_2} {k_3} \Big)\nonumber\\
   &&\Big(4 \Lambda +3 {k_1} ({k_2}+{k_3})-3 {k_2} {k_3} \Big)\,,
    \label{remainigcondition1}
    \eea
which allows for some particular solutions with $\alpha_2=0$ that we explore now.

\underline{\bf Some simple, particular solutions}

The last condition Eq. (\ref{remainigcondition}), linking
$k_1,k_2,k_3,\Lambda,\rho$, is quadratic in $\rho$.
There are particular cases
for which this quadratic term Eq. (\ref{remainigcondition1}) vanishes, thus making the solutions much simpler.
This happens for three special cases in which
$k_1,k_2,k_3,\Lambda$, are related according to:
\bea
4 \Lambda &=& 3 ({k_1}{k_2}-{k_1}{k_3}-{k_2}{k_3} )\,,\nonumber \\
4 \Lambda &=& 3 ({k_1}{k_3}-{k_1}{k_2}-{k_2}{k_3} )\,,\nonumber \\
4 \Lambda &=& 3 ({k_2}{k_3}-{k_1}{k_3}-{k_1}{k_2} )\,.
\label{speciallambdas}
\eea
All three cases are the same by exchanging indices, so we need only to examine the first of them.
Interestingly the solutions corresponding to the first case in Eq. (\ref{speciallambdas}) satisfy
$c_1=m k_1,\, c_2=m k_2$, with
$\displaystyle m=\frac12-\frac{s}{4}$ and $\displaystyle c_3=k_3\left(\frac{1}{4}+\frac{1}{s}\right)$.
where $\displaystyle s\equiv k_3\left(\frac{1}{k_1}+\frac{1}{k_2}\right)$\,.
These are exactly the solutions discussed in section \ref{GBd7}.

Another simple solution is obtained for the special case in which $k_1=k_2=k_3=k$.
Then it turns out
that $c_1=c_2=c_3=c$ with $\displaystyle c=\frac{27 k^2+4 \Lambda -2 \rho }{36 k}$,
whereas $k$ must satisfy
\beq
k= \frac{1}{9}(\Lambda + \rho)
\pm \frac{\sqrt{3}}{27} \sqrt{\rho ^2+14 \Lambda  \rho-5 \Lambda ^2}
\eeq
An equivalent way to express this solution is to take $k$ and $c$ as independent variables and
determine
$\Lambda,\, \rho$, through
$\displaystyle \Lambda=4\, c^2-\frac{3 }{4}k^2$ and
$\displaystyle \rho=8\, c^2-18\, c\, k+12\, k^2$.
This shows that for this particular solution
$\rho$ is also always positive.

Going back to the general solution of Eq. (\ref{linearsystem}), we may explore whether the
condition for this general solution
to be enlarged to higher dimensions $d>7$ with flat directions,
which is (see Appendix II) $\rho+4\Lambda=0$, can be satisfied in our case.
The answer is in the negative: the solutions become complex.


\section*{Acknowledgements}
Partial support for this work to AM was provided by FIS2015-65140-P (MIN\-E\-CO/FE\-DER). JMP acknowledges support by FPA2013-46570-C2-1-P, 2014-SGR-104 (Generalitat de Catalunya) and Consolider CPAN and by the Spanish goverment (MINECO/FEDER) under project MDM-2014-0369 of ICCUB (Unidad de Excelencia Mar\'\i a de Maeztu). 
ND thanks Albert Einstein Institute, Golm and University of Barcelona for visits
that have facilitated this work. 

\end{document}